\documentstyle[prcrc,11pt,fleqn]{article}

\title{Astrophysical search strategies for accelerator blind dark matter}

\author{James D. Wells\address{Stanford Linear Accelerator Center,
	Stanford University, Stanford, CA 94309}
	\thanks{Work supported by the Department of Energy, contract
         DE-AC03-76SF00515. SLAC-PUB-7807.}}

\begin{document}
\maketitle

\begin{abstract}
A weakly interacting dark-matter particle may be very difficult to
discover at an accelerator because it either (1) is too heavy, 
(2) has no standard-model gauge interactions, or (3) is almost degenerate
with other states.  In each of these cases, searches for annihilation
products in the Galactic halo are useful probes of dark-matter properties.
Using the example of supersymmetric dark matter, I demonstrate how 
astrophysical searches for dark matter may provide discovery and mass
information inaccessible
to collider physics programs such as the Tevatron and LHC.

\end{abstract}

\section{Introduction}

A stable, weakly interacting particle explanation for dark matter (DM) is
attractive~\cite{r1}. This is because the astrophysics community declares its 
favorability from galaxy rotation
curves, structure formation, etc., and the particle
physics community has recognized that the lightest supersymmetric 
(SUSY) particle
(LSP) is generically stable with ${\cal O}(\rho_c)$ relic abundance~\cite{r2}.
Rather
than causing a problem, the LSP 
provides a solution to the astrophysics concerns.
Although axions and other DM candidates can be made to fit the
data just as well as the LSP, it is perhaps a little less compelling since
arbitrary axion parameters do not yield viable DM.  This is purely
aesthetic, and only experimental probes are allowed to make these decisions.

The focus of this article is
the relationship between the SUSY theory
of dark matter and the experimental probes of it.  Often particle physicists
think of the large hadron collider (LHC) as a kind of death for good people:
when it happens we'll know all the answers.  It is true that the LHC will
have a tremendous mass reach for supersymmetry, and if nothing is found
then it will be an unpleasant few weeks for particle physicists.  
However, these two extremes of thinking are not likely to be relevant.  
More likely, we will experience with the Tevatron and LHC
a large collection of interesting 
observables that will be difficult even to interpret conclusively as
supersymmetry (or some other theory).  Not only that, even if some
chargeless DM particles were produced at the colliders,
we would only be able to say that the particles are
stable on time scales less than
the detector radius.  Experiments devoted to discovering and confirming
DM are necessary.

It could happen that the DM will not be seen at the colliders;
or, it is seen but it will be difficult to say what its mass is.  This is
addressed specifically in this paper.  There are probably
other reasons why this could happen, but in the SUSY framework
there are three good reasons: (1) The LSP is too heavy to be produced,
(2) the LSP has no standard-model gauge interactions, and (3) the LSP is
stable with other particles.

\section{Heavy dark matter}

In SUSY, the expectation is that the LSP is near the weak
scale.  This is because the same SUSY-breaking mass scale
that sets the LSP mass also sets the $W$ mass.  However,
one immediately encounters the question: is 200 GeV ``near the weak scale''?
Or, is 2 TeV ``near the weak scale''?  The question is morphed into
a response by devising a fine-tuning parameter that essentially
indicates how far above the weak scale one is allowed to go and still
call it ``near the weak scale.''  Again, the largest allowed mass scale
of the LSP is not a question that humans are supposed to sound confident
answering. 

A question that we can answer is how far above the weak scale
would SUSY have to be for us not to see it?  Given a 
well-formulated theory, we can analyze it and answer this question.  In 
so-called minimal supergravity scenarios with common scalar and common
gaugino masses, the answer is that the gluinos and squarks
have to be less than a few hundred GeV at the Tevatron and
2~TeV at the LHC~\cite{r3}.  Using renormalization group relations
that predict the LSP mass in terms of these masses, we can
conclude that the mass of the LSP must be below about 150 GeV to be
seen at the Tevatron and perhaps 350 GeV to be seen at the LHC.

In minimal supergravity theories, the bino (superpartner of the hypercharge
gauge boson) is generally the LSP~\cite{r2}.  It has some small mixing with the
superpartners of the Higgs boson, which can play an interesting role
in some observables (especially LSP scattering off nuclei in 
cryogenic detectors).  For accelerator physics and annihilations in
the Galactic halo,
the bino component of the LSP is usually most important.  

Since the LSP is a Majorana particle, annihilations into final state fermions
must flip chirality in the $S$-wave.  For $m_{\rm LSP}< m_t$ this chirality
flip is highly suppressing:
$(\sigma v)_{S} \propto m^2_f/\tilde m^4$.
Therefore, the annihilation proceeds through a $P$-wave, which is
velocity suppressed (the LSP is a ``cold relic'' with non-relativistic
energies). However, the annihilation rate in the galactic halo must
proceed through the $S$-wave since the virial velocity today of the 
LSPs is only a few hundred kilometers per second (highly non-relativistic).
So, it becomes a little tricky to correlate the relic abundance of a
particle with its annihilation rate in the Galactic halo.

When the LSP is much heavier, this correlation becomes easier.
The relic abundance now has a potentially large $S$-wave diagram proportional
to $m_t^2$, and annihilations of the LSP in the Galactic halo do the same.
Therefore,
a one-to-one correspondence can be written for the two.  Since binos cannot
couple to winos or vector bosons, heavy LSP
dark matter will want to annihilate almost 100\% of the time into
top quark pairs.  One of the best DM observables~\cite{r4} for this
annihilation arises from $\chi\chi\to t\bar t$, where $t\to bW^+$
and then $W^+\to e^+\nu$.  The positrons can have an interesting
energy profile from this annihilation signal. When contrasted with the
energy profile of positrons from ordinary QED processes, a bump or shoulder
is expected in the spectrum.  

For LSP annihilations near the threshold of top pair production,
there is a higher $e^+$ peak (near $30$~GeV) 
corresponding to $W^+\to e^+\nu$ decays and a
lower peak corresponding 
to $b\to e^+\nu c$ decays in the $t\bar t$ events.  Positrons
from fragmentation of jets also contribute a continuum spectrum at
the lower energies. These continuum positrons are difficult to separate
from background positrons.  The all-electron spectrum measurements 
appeared to be slightly peaking
in the $\sim 30$~GeV region, although
the most recent and precise measurements are not conclusive~\cite{r5}.  

\section{Dark matter with no gauge interactions}

It is also possible that the DM has no gauge interactions allowed.
In the case of the bino, since it is the superpartner of the hypercharge
gauge boson, one expects it to interact by gauge interactions with
the right-handed sleptons, for example.  Indeed, it is 
these gauge interactions that set the relic abundance of the bino LSP.
However, if the dark matter is the superpartner of a Higgs singlet,
then it has no gauge interactions at all, and cascade decays of MSSM
states may not terminate with the true LSP inside the detector volume.

The superpotential of a singlet Higgs SUSY theory contains the
terms $W=\lambda SH_u\cdot H_d+{\lambda'}{S^3}/3$.
The fermionic component, $\chi_S$, 
of the $S$ chiral superfield could be the dark
matter and it could annihilate into Higgs bosons if heavy enough.
This possibility does not preclude interesting studies at colliders;
however, I have separated it out as a good theory for astrophysical searches
for two reasons.  (1) In order for $\chi_S$ to be a good DM
candidate it must be fairly heavy in order to annihilate into, for example,
$h^0+A^0$ final states. (2) These
theories could have a significantly larger monochromatic two-photon signature
from annihilations in the Galactic halo compared to ordinary minimal 
supergravity models.

The annihilation of $\chi_S\chi_S\to \gamma\gamma$ can occur via
a pure Higgsino internal loop of particles enhanced by $\lambda^4$, if
$\lambda$ is rather large.  Even if it is 1, the enhancement over
minimal supergravity models is at least as high as $g_1^{-2}\sim 10$.
One can compare the scatter-plot 
points in P.~Ullio's results~\cite{r6} for monochromatic photon
flux and multiply by roughly
an order of magnitude for the highest flux models at a given LSP mass
and estimate the $\chi_S\chi_S$ signal.  A long-exposure
GLAST-like detector~\cite{r7} with high energy resolution would be ideal to
measure this signal.

\section{Dark matter mass-degenerate with other particles}

If the dark matter is degenerate with other particles, it might
be difficult to find any particle.  This is the case with the Higgsino
LSP.  The LSP is a singlet state of the Higgsinos and there is a triplet
multiplet of Higgsinos just above the LSP.  In collider experiments one
often relies on the leptons from cascades of the next heaviest chargino
or neutralino into the lightest neutralino (LSP).  If there is mass
degeneracy between these states, then the leptons will be very soft and
undetected.  

Astrophysical searches are good probes of the Higgsino LSP.  The monochromatic
two-photon searches~\cite{r6} are especially useful, since the signal
is expected to be rather large.
In addition
to the excitement that would arise by seeing such a signal, it could
provide mass resolution that the Tevatron and LHC just could not provide.
The GLAST detector, for example, could resolve an $\sim 100$~GeV dark-matter
peak on the order of a percent or two mass resolution~\cite{r7}.  
Tevatron and LHC
have no absolute scale capabilities to measure the mass of the LSP, but rather
can do fairly well with mass differences. For example in the 
decay $\chi^0_2\to \mu^+\mu^-\chi^0_1$, the invariant mass distribution
of the muons can tell us the mass difference between $\chi^0_2$
and $\chi^0_1$ (the LSP). The absolute mass scale is difficult to extract
in a general approach to LHC observables.  However, the two photon peak
can tell us this number to within a few percent.

\section{Conclusion}

So far, the usefulness of the $\bar p$-searches for
DM has not been discussed.  
This is a very unique search strategy, because the
signal is never expected to have any energy peaking associated with it.
In the case of the positron and photon searches for DM, the energy
peaks were necessary to resolve the signal from background.
In the $\bar p$ observables, it is the {\it background} that has an
energy peak.  The secondary $\bar p$ flux from spallation peaks at
about $1$~GeV. This is easily derived from maintaining Lorentz invariance
and baryon number conservation in $pp$ collisions.  The interstellar
$\bar p$ spectrum quickly falls above and below 1~GeV~\cite{r8}.  
Above 1~GeV the
SUSY prediction falls rapidly as well, and so it is not as
useful; however, below 1~GeV the supersymmetric LSP annihilations can
produce a large interstellar $\bar p$ flux measurable above the background.

The challenge with antiproton searches is solar modulation.  The solar wind
slows low-energy protons and antiprotons~\cite{r8}.
Thus when a proton or antiproton with kinetic energy less than 1~GeV
enters
the heliosphere, it might not be able to ``swim upstream'' to the
earth-based detector, and if it does the energy could be drastically changed.
Sophisticated modelling exists for these complicated effects, but it might 
be difficult to have
confidence in a signal. For this reason,
it could be useful to put an antiproton spectrometer on the recently
considered interstellar probe~\cite{r9}.  
It might take a few decades to reach beyond
the $\sim 100$~AU required to get unambiguous results, but it is a relatively
inexpensive piggy-back payload that has potentially enormous 
payoffs~\cite{r10}.
For example, primordial black holes, which (hopefully) have no chance of 
being produced at a collider experiment can evaporate antiprotons at a
significant rate.  Probing their existence is perhaps best accomplished
with an interstellar antiproton spectrometer.

\end{document}